\documentclass[useAMS,usegraphicx,usenatbib]{mn2e}
\usepackage{times}

\title[Optical Variability and Colour Behaviour of 3C~345]
      {Optical Variability and Colour Behaviour of 3C~345}

\author[J. Wu et al.]{Jianghua Wu$^1$\thanks{E-mail: jhwu@nao.cas.cn},
       Xu Zhou$^1$, Jun Ma$^1$ and Zhaoji Jiang$^1$ \\
   $^1$Key Laboratory of Optical Astronomy, National Astronomical
       Observatories, Chinese Academy of Sciences, \\
       20A Datun Road, Chaoyang District, Beijing 100012, China}

\begin{document}

\date{}

\maketitle

\begin{abstract}
The colour behaviour of blazars is a subject of much debate. One argument
is that the BL Lac objects show bluer-when-brighter chromatism while the
flat-spectrum radio quasars (FSRQs) display redder-when-brighter trend. Base
on a 3.5-year three-colour monitoring programme, we studied the optical
variability and colour behaviour of one FSRQ, 3C 345. There is at least one
outburst in this period. The overall variation amplitude is 2.640 mags in the
$i$ band. Intra-night variability was observed on two nights. The
bluer-when-brighter and redder-when-brighter chromatisms were simultaneously
observed in this object when using different pairs of passbands to compute
the colours. The bluer-when-brighter chromatism is a shared property with the
BL Lacs, while the redder-when-brighter trend is likely due to two less
variable emission features, the Mg\,{\sc ii} line and the blue bump, at short
wavelengths. With numerical simulations, we show that some other strong but
less variable emission lines in the spectrum of FSRQs may also significantly
alter their colour behaviour. Then the colour behaviour of an FSRQ is linked
not only to the emission process in the relativistic jet, but also to the
redshift, the passbands used for computing the colour and the strengths of
the less variable emission features relative to the strength of the
non-thermal continuum.
\end{abstract}

\begin{keywords}
galaxies: active --- quasars: individual (3C~345).
\end{keywords}

\section{Introduction}
Blazars are the most variable subclass of active galactic nuclei (AGNs). The
most prominent property of blazars is the rapid and strong variability in
their continuum emission. The continuum emission and its variability is
believed to originate from the relativistic jet pointed basically to our
line of sight. In the frame work of leptonic model, the low energy continuum
emission of blazars has a synchrotron origin, while the high energy continuum
emission is from inverse Compton upscattering of the low energy emission by
the relativistic electrons in the jet. The low energy seed photons may either
be produced in the jet (the synchrotron self-Compton or SSC process) or come
from the accretion disc, broad line region (BLR), or dusty torus \citep[for a
recent review, see][]{bottcher07a}. Blazars can be subdivided into
flat-spectrum radio quasars (FSRQs) and BL Lacs, depending on whether or
not there are strong emission lines in their spectra.

3C~345 is the first established variable quasar \citep{burbidge65} and is
classified as FSRQs later. Its redshift is 0.5928 \citep{marz96} and its
position is 16:42:58.8, 39:48:37 (J2000.0). The high declination enables
short monitoring gap ($\sim$ 80 days) each year. Ever since its discovery,
it has been monitored intensively and a wealth of data have been collected.
Its optical flux shows rapid variations occurring in a few days or weeks
superimposed upon a slowly varying component \citep{mcgimsey75}. Several
large-amplitude outbursts ($\ge1.5$ mags) were observed
\citep[e.g.][]{schramm93,webb94}. Based on a collection of the historical
data and on their own observations, \citet{belokon99} found that 3C~345
varied by more than 3.0 mags in the $B$ band from 1965 to 1995. On short
time-scales, \citet{baba85} reported a brightening of 0.48 $B$ mags in just
half an hour on JD 245 4230. \citet{kidger90a} even claimed a 0.47 $B$ mags
brightness drop in 13 minutes. Despite these rapid variations, \citet{mihov08}
did not find significant intra-night optical variability (INOV) in this object.
The optical polarization of 3C~345 is highly variable between
5\% and 35\% and is strongly correlated with brightness and wavelength
\citep{smith86}. 

In the radio domain, several large amplitude outbursts were observed in this
object (see the website of Radio Astronomical Observatory, University of
Michigan, http://www.astro.lsa.umich.edu/obs/radiotel/umrao.php). The
outbursts occurred at high frequency and propagated gradually to lower
frequencies with gradually decreasing amplitudes \citep{aller85,bregman86,
webb94}. The high-frequency ($\ge 14$ GHz) radio flares occurred almost
simultaneously with the optical ones, while the lower-frequency (4.8 and
8 GHz) flares lagged the optical-infrared flares by roughly 1 years
\citep{bregman86,webb94,lobanov05}. Superluminal components were identified
in the jet of this object \citep*{perley82,kollgaard89}.

The period was searched in the variability of 3C~345 since late 1960's.
The claimed periods range from 80 days \citep{kinman68} to 11.4 years
\citep{webb88}. However, some more recent results didn't show any period
\citep[e.g.,][]{kidger86,kidger89,schramm93}. A few models were proposed to
explain the quasi-periodic variability of 3C~345, such as the lighthouse
model by \citet{schramm93} and the binary black hole models by
\citet{caproni04} and by \citet{lobanov05}.

The colour behaviour of blazars is a subject of much debate. Some authors
found a bluer-when-brighter (BWB) chromatism \citep[e.g.,][]{vagnetti03,
wu05,wu07}, some others claimed the opposite, namely, a redder-when-brighter
(RWB) trend \citep[e.g.,][]{ramirez04}, or no clear tendency \citep[e.g.,][]
{bottcher07b,bottcher09}. The same object may show different trends in
different variation modes \citep [e.g.,][]{poon09} or on different time-scales
\citep[e.g.,][]{ghise97,raiteri03}. Several authors argued that BL Lacs
display BWB chromatism while FSRQs show RWB trend \citep{fan00,gu06,hu06,
rani10}.

We have monitored 3C~345 since 2006. More than 700 data points were
collected on 53 nights. The long- and short-term variability of this object
was studied and the colour behaviour was investigated. Here we present the
results.

\begin{figure}
\includegraphics[width=8.4cm]{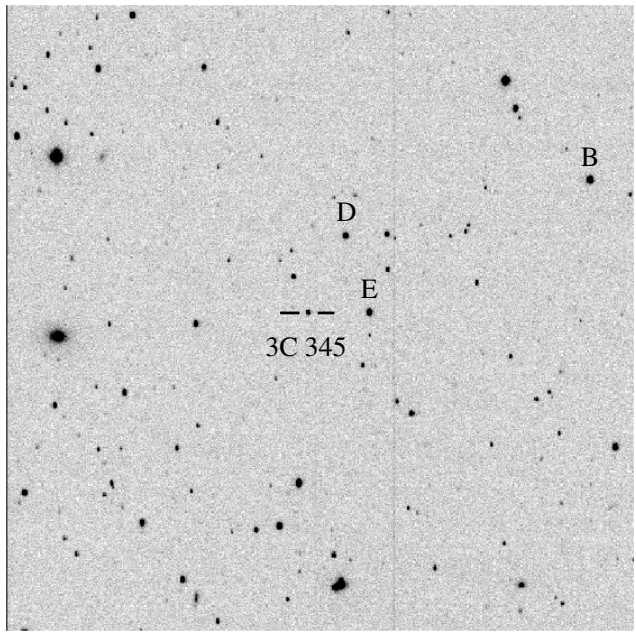}
\caption{Finding chart of 3C~345, taken in the $i$ band on JD 245 4954. The
blazar and three comparison stars are labelled. North is to the up and east,
to the left.}
\label{F1}
\end{figure}

\section{Observations and Data Reductions}
The monitoring was performed with a 60/90 cm Schmidt telescope at Xinglong
station, National Astronomical Observatories of China. The telescope is
equipped with a $4096\times4096$ E2V CCD and 15 intermediate-band filters,
which are used to do the Beijing-Arizona-Taiwan-Connecticut (BATC) survey
\citep{zhou05}. The CCD has a pixel size of
12 $\micron$ and a spatial resolution of 1.3 $\arcsec{\rm pixel}^{-1}$. When
used for blazar monitoring, only the central $512\times512$ pixels are read
out as a frame in order to reduce the readout time and to increase the
sampling rate. Each such frame has a field of view of about $11'\times11'$.
The monitoring of 3C~345 was made in five intermediate bands, $c$, $e$, $i$,
 $m$ and $o$ ($e$, $i$ and $m$ bands in 2006 and $c$, $i$ and $o$ bands
from 2007 to 2009). Their central wavelengths and bandwidths are listed in
Table~1. The exposure times are generally 540, 300, 180, 300 and 480 s in
the $c$, $e$, $i$, $m$ and $o$ bands, respectively, and may vary slightly
depending on the weather and moon phase. An example frame in the $i$ band is
shown in Fig.~\ref{F1}. 3C~345 and three reference stars are labelled. The
reference stars are adopted from \citet{smith85}. Their $c$, $e$, $i$, $m$
and $o$ magnitudes were obtained with observations on several photometric
nights and are presented in Table~2. The data presented here
cover the period from 2006 February 16 (JD 245 3783) to 2009 June 1 (JD 245
4984). More than 700 data points were collected on 53 nights.

\begin{table}
\caption{Central Wavelengths and Bandwidths of 5 BATC Filters}
\begin{tabular}{@{}ccc}
\hline
Filter & Central Wavelength & Bandwidth \\
       & (\AA) & (\AA) \\
\hline
 $c$ & 4206 & 289 \\
 $e$ & 4885 & 372 \\
 $i$ & 6685 & 514 \\
 $m$ & 8013 & 287 \\
 $o$ & 9173 & 248 \\
\hline
\end{tabular}
\end{table}

The data reduction procedures include bias subtraction, flat-fielding,
extraction of instrumental aperture magnitude, and flux calibration. The
radii of the aperture and the sky annuli were adopted as 3, 7 and 10 pixels,
respectively. The brightness of 3C~345 was calibrated relative to the average
brightness of stars B and D. Star E acts as a check star. Its differential
magnitude was also calculated relative to the average brightness of stars B
and D, and was used to verify the accuracy of our observations.

\begin{table}
\caption{BATC $c$, $e$, $i$, $m$ and $o$ magnitudes of reference stars}
\begin{tabular}{@{}cccc}
\hline
Passband & B & D & E \\
\hline
$c$ & 14.933 & 16.286 & 16.951 \\
$e$ & 14.782 & 15.885 & 16.183 \\
$i$ & 14.118 & 15.069 & 14.728 \\
$m$ & 13.974 & 14.886 & 14.332 \\
$o$ & 13.852 & 14.706 & 14.115 \\
\hline
\end{tabular}
\end{table}

\begin{table}
\caption{Statics on INOVs}
\begin{tabular}{@{}ccccc}
\hline
JD & Band & Duration & $C$ & Amplitude \\
   &      & (hour) &  &  (mag) \\
\hline
245 3783 & $e$ & 1.62 & 2.750 & 0.104 \\
245 3783 & $i$ & 1.62 & 8.286 & 0.153 \\
245 3783 & $m$ & 1.62 & 5.545 & 0.154 \\
245 3786 & $i$ & 2.24 & 7.889 & 0.437 \\
245 3786 & $m$ & 2.25 & 5.048 & 0.252 \\
\hline
\end{tabular}
\end{table}

\section{Light Curves and Variation Amplitudes}
\begin{figure}
\includegraphics[width=8.4cm]{fig/lc3783e.ps}
\includegraphics[width=8.4cm]{fig/lc3783i.ps}
\includegraphics[width=8.4cm]{fig/lc3783m.ps}
\caption{Intra-night light curves of 3C~345 in the $e$, $i$ and $m$ bands
on JD 245 3783 (large panels) and of the check stars (small panels).}
\label{F2}
\end{figure}

\begin{figure}
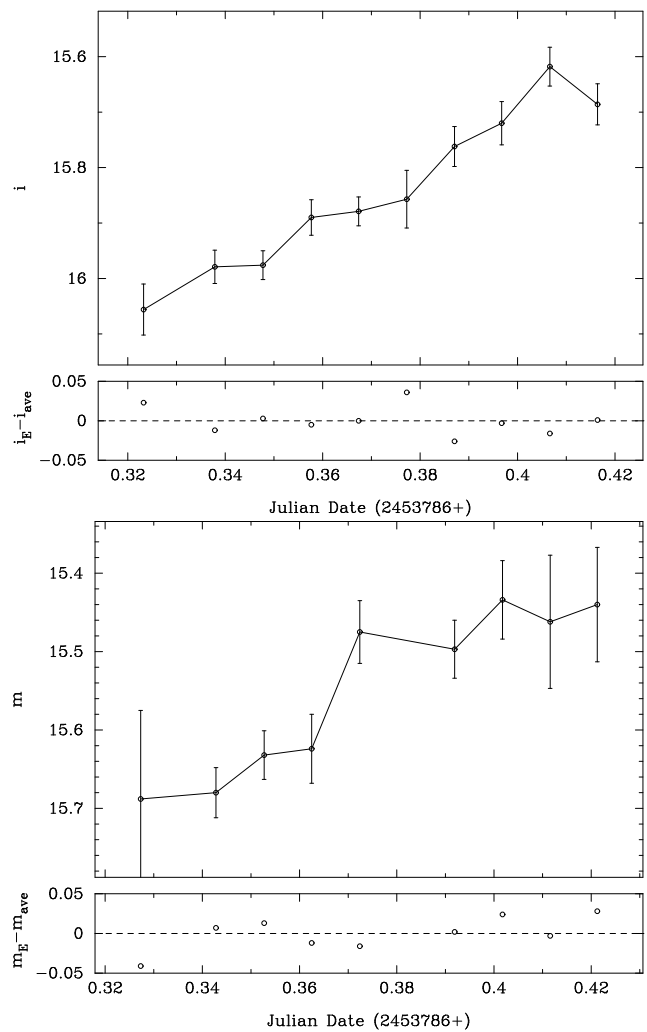

\includegraphics[width=8.4cm]{fig/lc3786i.ps}
\includegraphics[width=8.4cm]{fig/lc3786m.ps}
\caption{Intra-night light curves of 3C~345 in the $i$ and $m$ bands on
JD 245 3786 (large panels) and of the check stars (small panels).}
\label{F3}
\end{figure}

Among the 53 nights, 3C~345 shows INOVs on only 2 nights
in 2006. The light curves on these 2 nights are displayed in Figs.~\ref{F2}
and \ref{F3}. The large panels show the light curves of 3C~345, while the
small panels give those of the check star. Following \citet{jang97} and \citet*
{romero99}, a variability parameter is defined as $C=\sigma_{\rmn T}/\sigma$,
where $\sigma_{\rmn T}$ is the standard deviation of the magnitudes of the
target blazar and $\sigma$ is that of the check star. The latter can be taken
as the typical measurement error on a certain night.  When $C\geqslant2.576$,
the object can be claimed to be variable at the 99\% confidence level. The
variability parameters were calculated for the five intra-night light curves,
and the results are listed in Table~3. All five $C$ values are greater than
2.576, thus confirming the INOVs on these two nights. The variation
amplitudes, as defined by \citet{heidt96}, are also given in Table~3.
The amplitude in the $e$ band is much less than those in the $i$ and $m$
bands on the same night.

\begin{figure*}
\includegraphics[width=17.5cm]{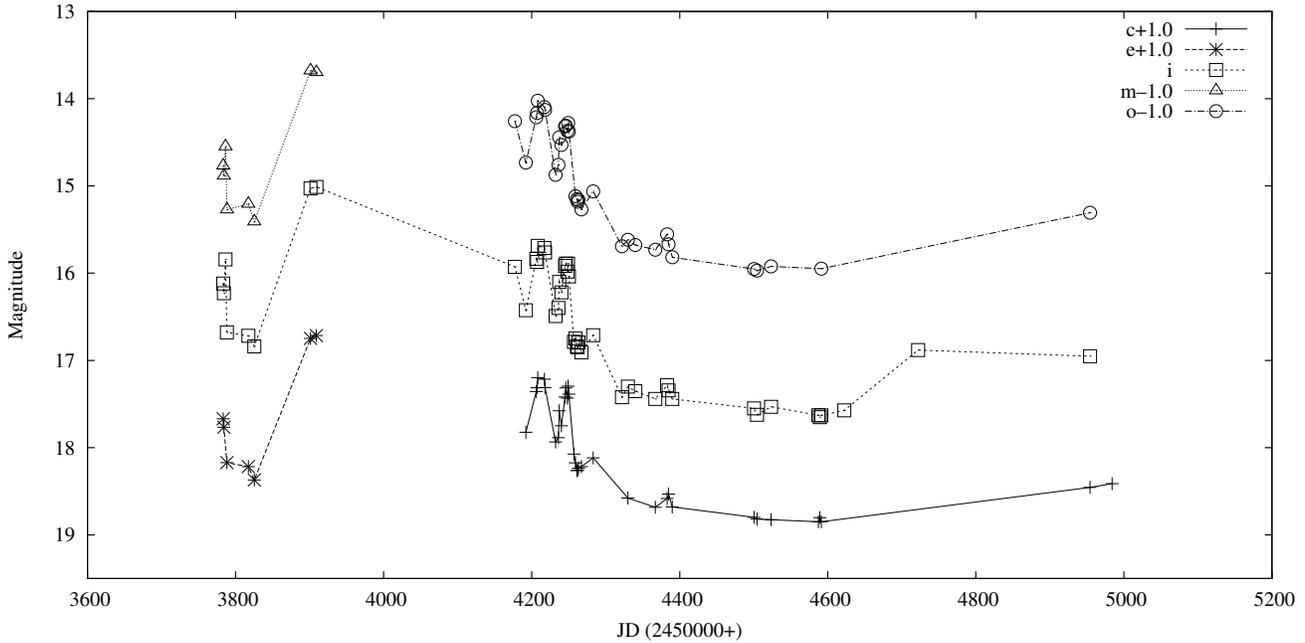}
\caption{Nightly-mean night curves of 3C~345 in five bands. For clarity, the
$c$-, $e$-, $m$- and $o$-band magnitudes are shifted by 1.0, 1.0, $-$1.0 and
$-$1.0 magnitudes, respectively.}
\label{F4}
\end{figure*}

3C~345 did not show INOV on the remaining 51 nights. Part of the reason
may be the relatively short monitoring period on each single
night in 2007--2009. Then its nightly average magnitudes were calculated
and used for the following analyses. The nightly-averaged light curves are
plotted in Fig.~\ref{F4}. Data in 5 different passbands are denoted with
different symbols and are connected with different lines.
One can see that there is at least one outburst from 2006 (JD 245 3800) to
the first half of 2007 (JD 245 4250). Our recorded peak is at JD 245 3909,
with $i=15.011\pm0.011$, which corresponds to $R\sim15.008$ with a simple
interpolation between the fluxes in the $e$ and $i$ bands. However,
we have very few observations during this period. So we cannot identify the
exact time(s) and amplitude(s) (and/or number) of the outburst(s). From around
JD 245 4200 to JD 245 4600, the object was monitored more or less constantly
and declined gradually in brightness. After that, we again have few data
and the object showed a tendency to recover till the end of our monitoring.
The overall amplitude in the $i$ band (used in the whole monitoring period)
is 2.640 mags. In 2006, where the source was monitored in the $e$, $i$ and
$m$ bands, the variation amplitudes in the three bands are respectively 1.656,
1.829 and 1.731 mags. In 2007--2009 (or from JD 245 4192 to JD 245 4591),
where the source was monitored in the $c$, $i$ and $o$ bands, the amplitudes
are respectively 1.652, 1.964 and 1.947 mags. Then two conclusions can be
drawn. Firstly, the amplitude in the $e$ or $c$ band is the smallest among
those in the three bands ($e$, $i$ and $m$, or $c$, $i$ and $o$). This is
similar to the case of the INOVs in this object mentioned above. Secondly, the
$i$ band amplitude is greater than the corresponding $m$ or $o$ band amplitude.

\begin{figure*}
\begin{minipage}{17.6cm}
\mbox{\includegraphics[height=6.1cm]{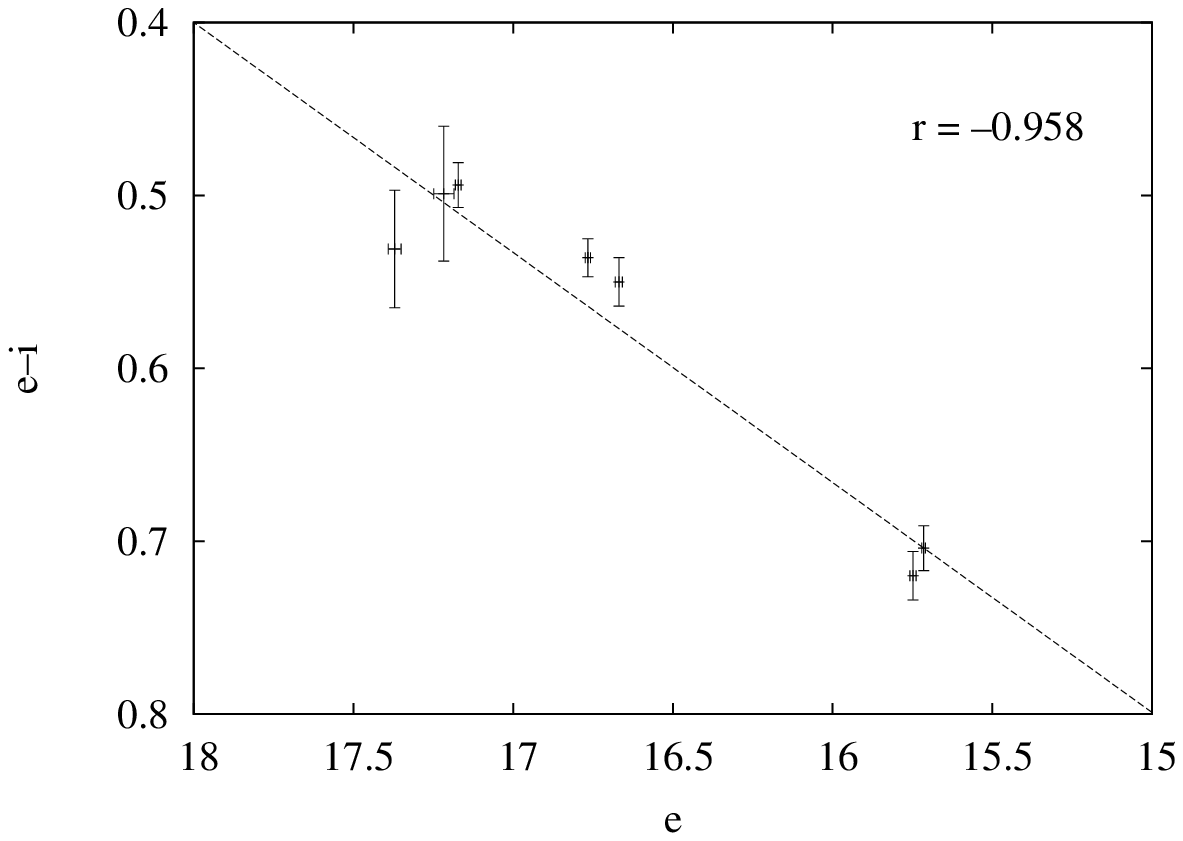}
      \includegraphics[height=6.1cm]{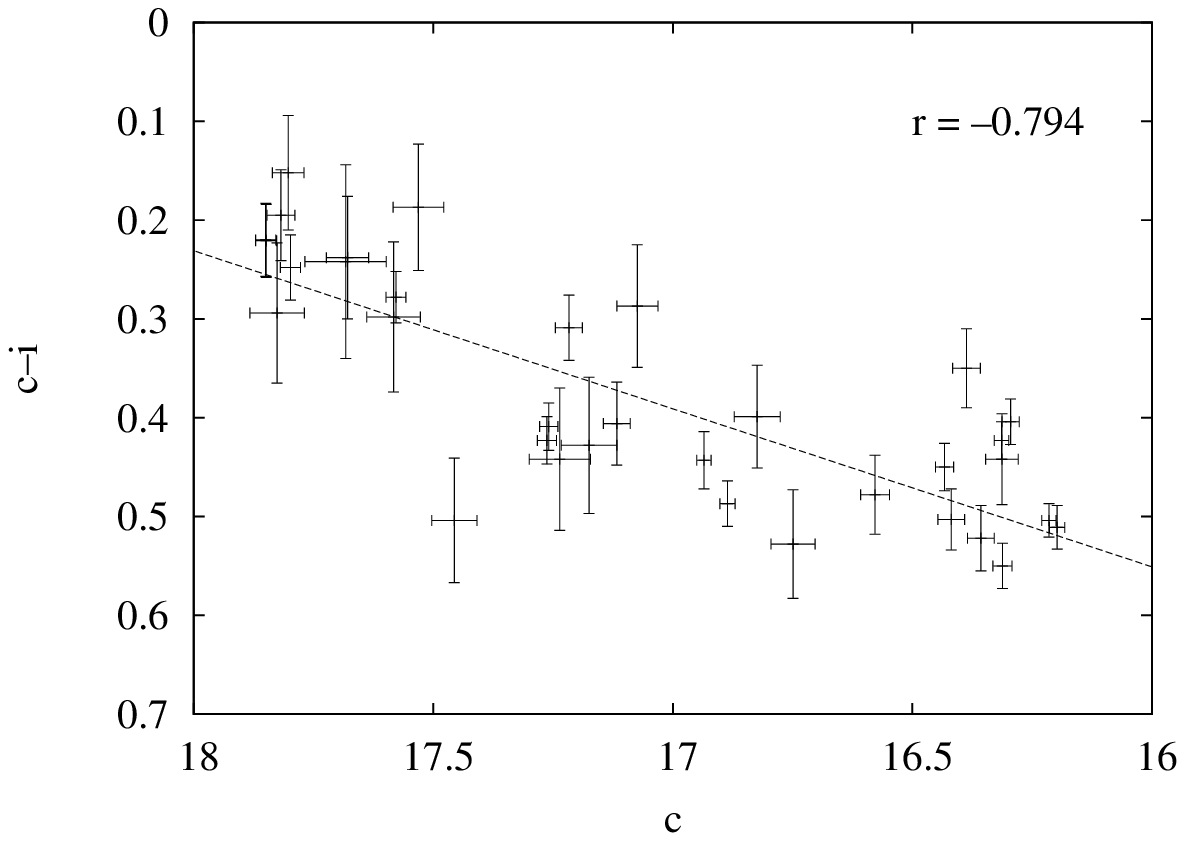}}
\mbox{\includegraphics[height=6.1cm]{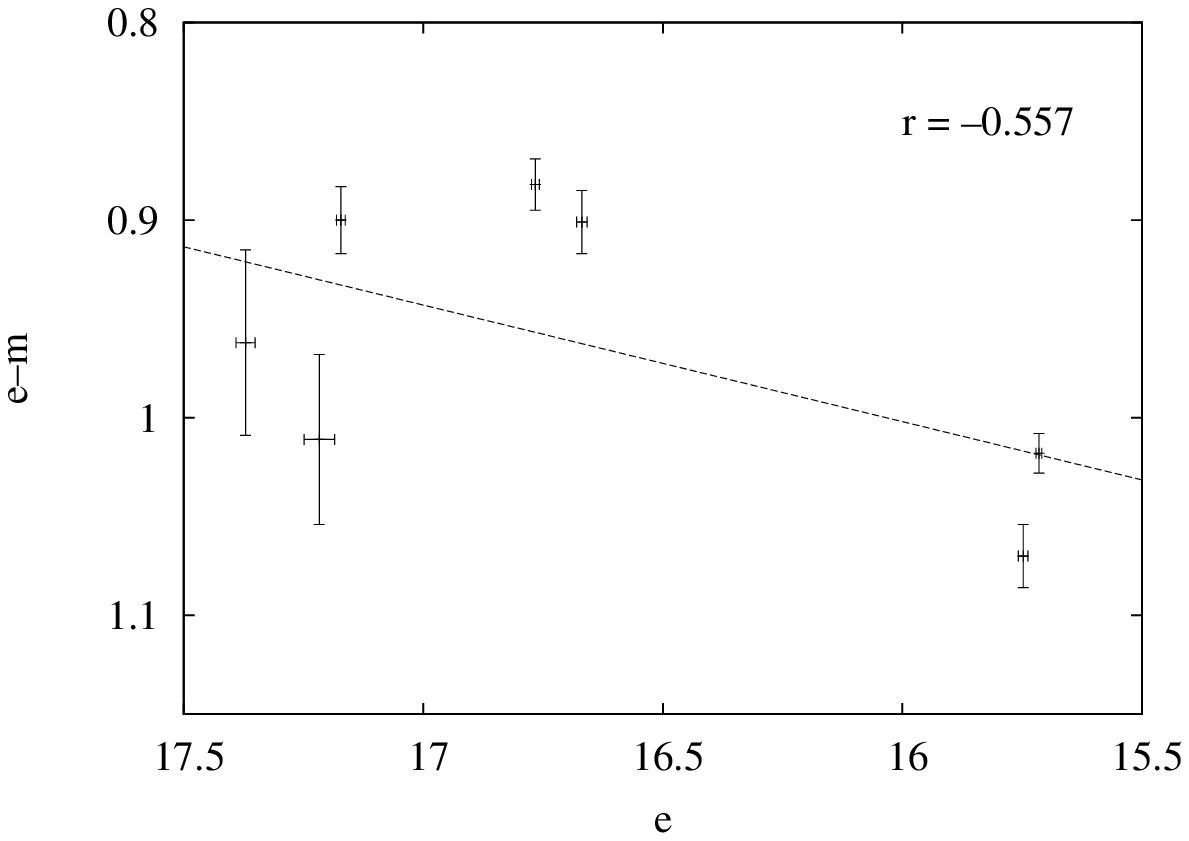}
      \includegraphics[height=6.1cm]{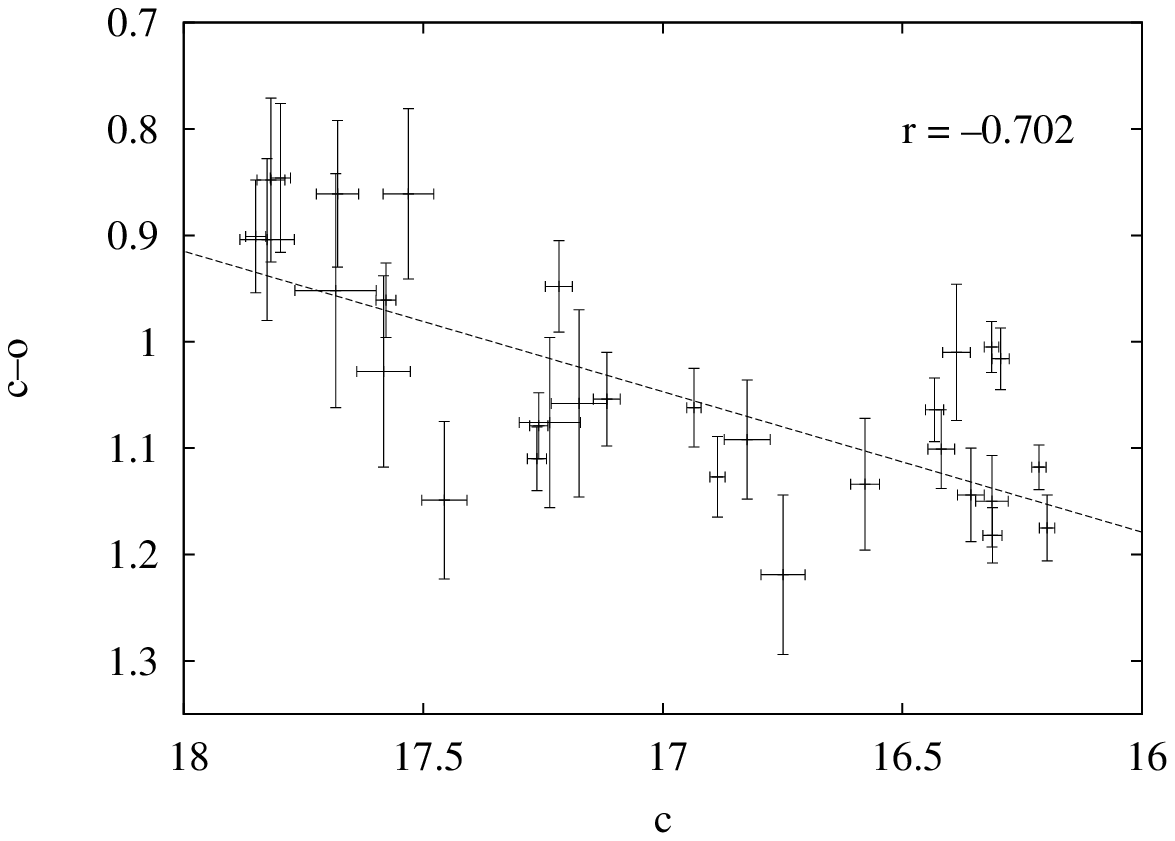}}
\mbox{\includegraphics[height=6.1cm]{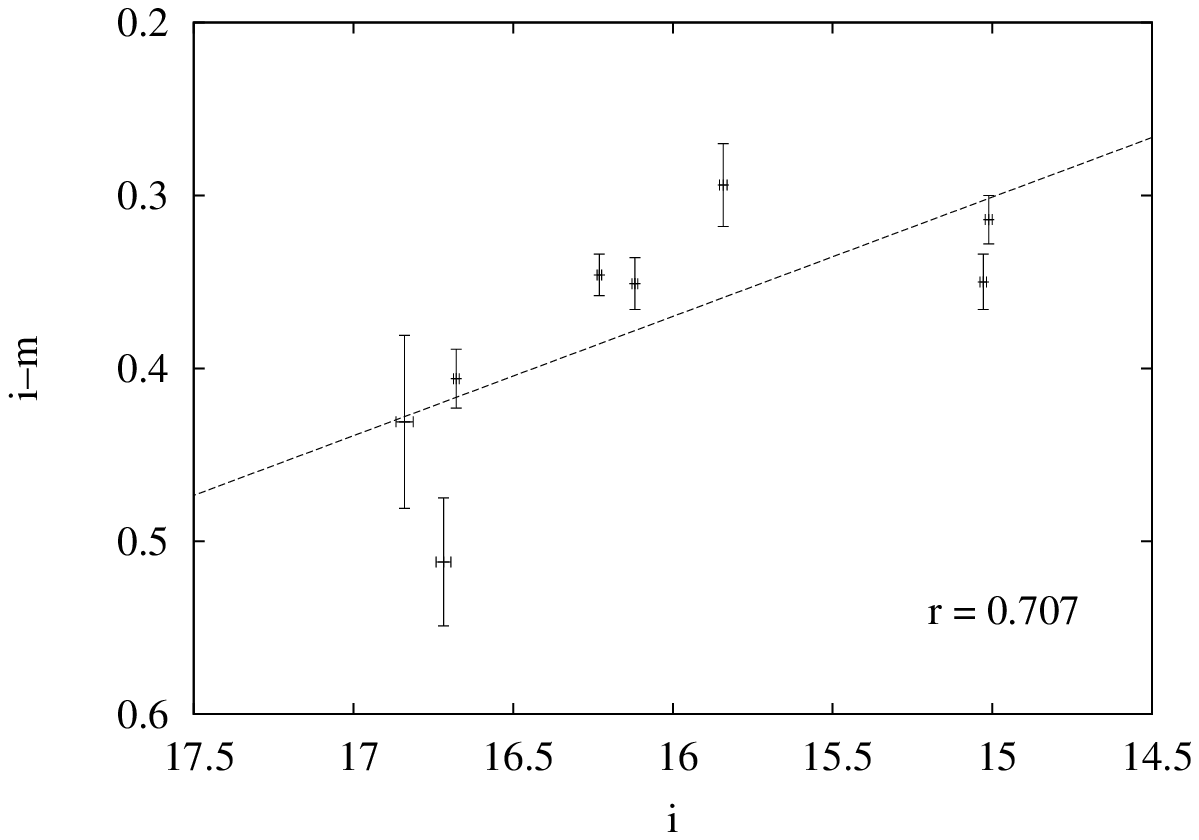}
      \includegraphics[height=6.1cm]{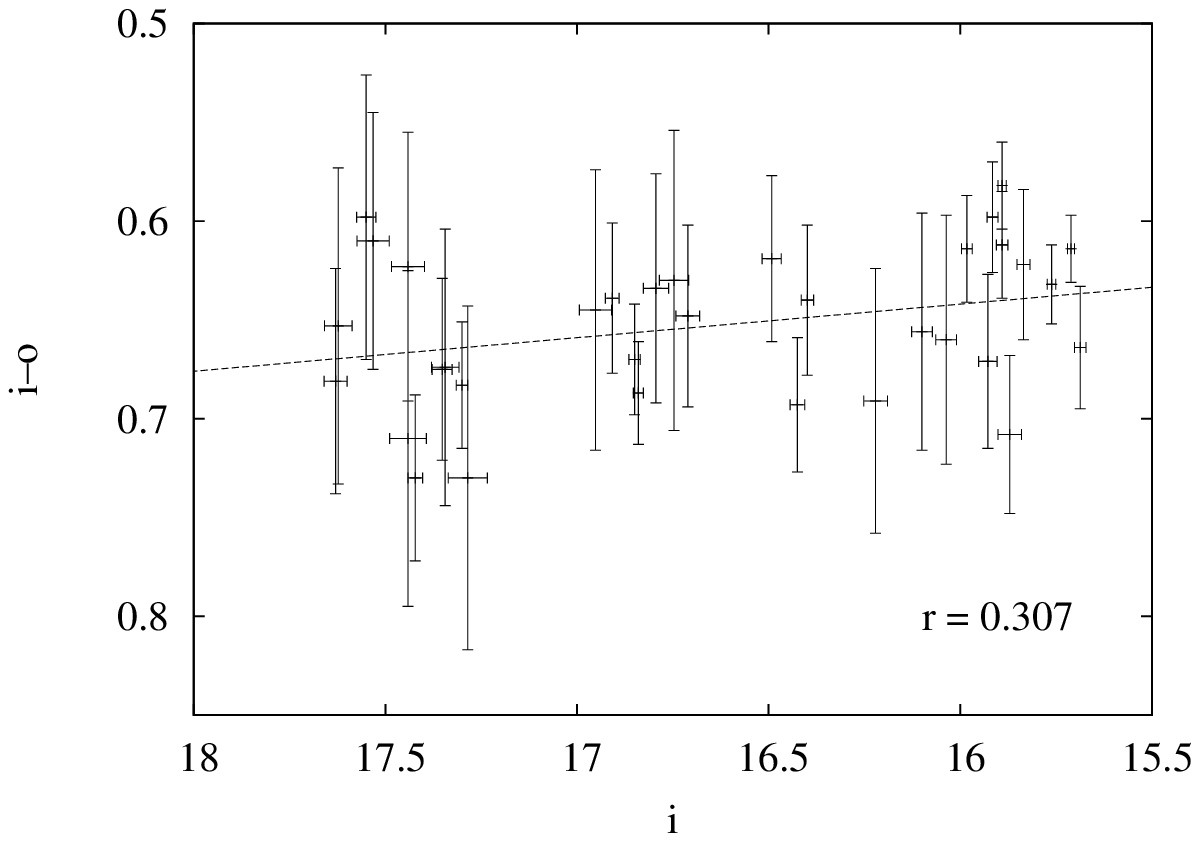}}
\caption{Colour-magnitude diagrams of 3C~345 in 2006 (left panels) and in
2007--2009 (right panels). The upper four panels show RWB chromatism while
the bottom two panels show BWB chromatism.}
\label{F5}
\end{minipage}
\end{figure*}

\section{Colour Behaviour}
The long-term colour behaviours of 3C~345 were studied by using the
nightly-average magnitudes. For the 2006 data, the colour indices of $e-i$,
$e-m$ and $i-m$ were calculated and plotted against $e$, $e$ and $i$,
respectively, in the left panels of Fig.~\ref{F5}. For the 2007--2009 data,
the colour indices of $c-i$, $c-o$ and $i-o$ were calculated and plotted
against $c$, $c$ and $i$, respectively, in the right panels of Fig.~
\ref{F5}. The $e-i$, $e-m$, $c-i$ and $c-o$ colours got redder when the source
became brighter, whereas the $i-m$ and $i-o$ colours got bluer when the source
became brighter. The correlation coefficients are respectively $-$0.958,
$-$0.557, $-$0.794, $-$0.702, 0.707 and 0.307, and the chance probabilities
are respectively $6.83\times10^{-4}$, $1.94\times10^{-1}$, $2.16\times10^{-8}$,
$1.06\times10^{-5}$, $4.98\times10^{-2}$ and $7.72\times10^{-2}$. There are
much fewer data points in the left panels than in the right ones, so the
correlations in the left ones are weaker than those in the right ones, but
the overall trends are the same.

\begin{figure}
\includegraphics[width=8.4cm]{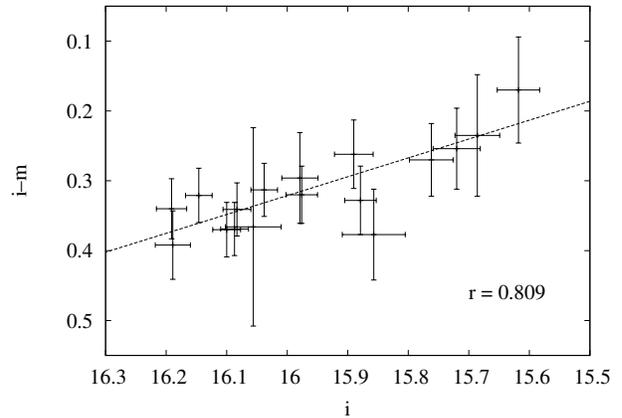}
\caption{Colour-magnitude diagram of 3C~345 on JDs 245 3783 and 245 3786,
the two nights showing INOVs. A strong BWB chromatism is visible.}
\label{F6}
\end{figure}

For the two nights (JDs 245 3783 and 245 3786) showing INOVs, the
colour-magnitude relation was also explored. Because there is no $e$-band
observation on the latter night, we only studied the $i-m$ colour versus
$i$ magnitude correlation. When calculating the $i-m$ colour, the $m$-band
light curve on the two nights were linearly interpolated or extrapolated
so as to get the $m$ magnitudes at exactly the same time when the $i$-band
observations were made. The result is displayed in Fig.~\ref{F6}. The
object tends to be BWB. The correlation coefficient is 0.809 and the chance
probability is $8.37\times10^{-5}$, indicating a strong BWB chromatism.

Therefore, both RWB and BWB chromatisms were observed in the variability of
3C~345. This is quite different from the previous results on the spectral or
colour behaviour of blazars, in which usually only one kind of colour behaviour
is reported for one object. Or at most, an object, except for displaying
either RWB or BWB chromatism in its middle- and short-term variability,
may be achromatic for its long-term variability \citep[e.g.,][]{ghise97,
villata02,raiteri03} or during a certain episode of time \citep*{wu05,poon09}.
One exception is that \citet{raiteri03} have recognized both RWB and BWB
trends in S5~0716+714, but at different times. Our study, however,
revealed that the RWB and BWB chromatisms appeared in 3C~345 at the same
time and/or in both the long-term variability and INOVs. This is the first
report of such a phenomenon in the study of the colour behaviour of blazars. 

\section{Discussions}
\begin{figure*}
\begin{minipage}{17.6cm}
\mbox{\includegraphics[height=6.1cm]{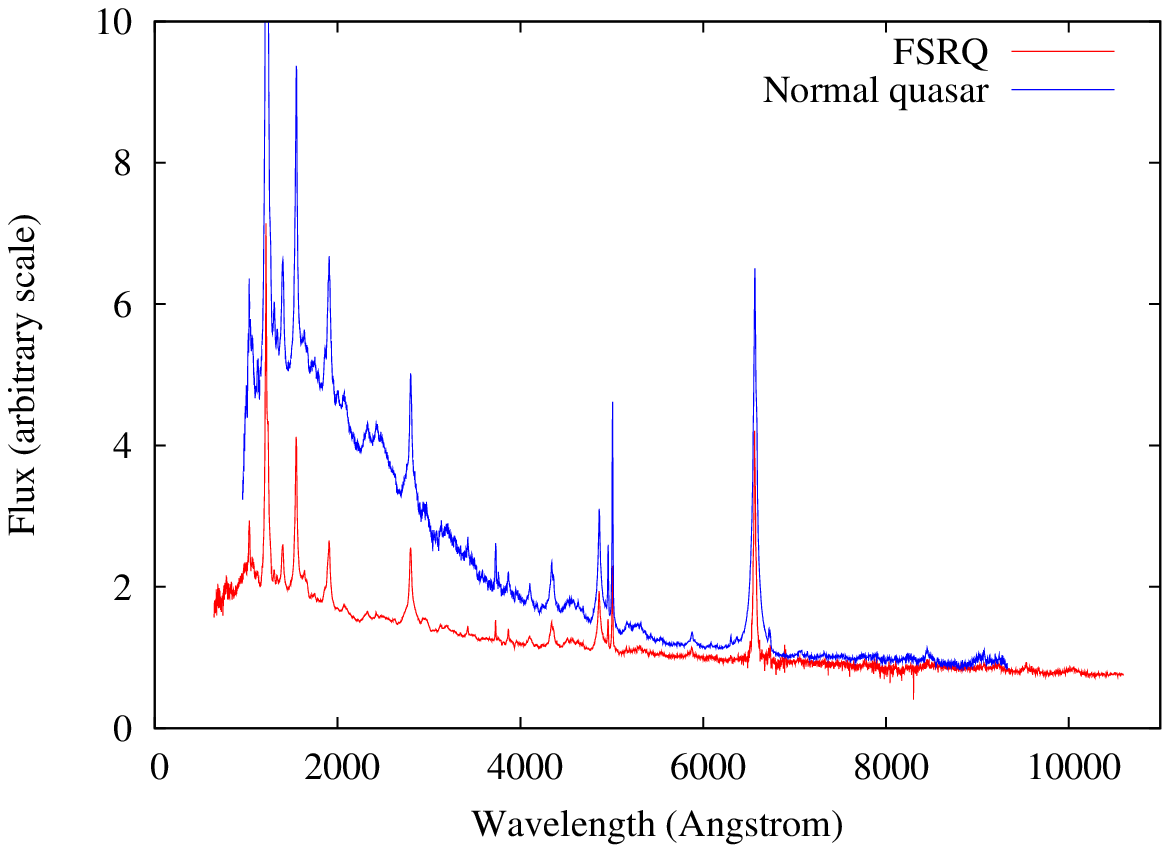}
      \includegraphics[height=6.1cm]{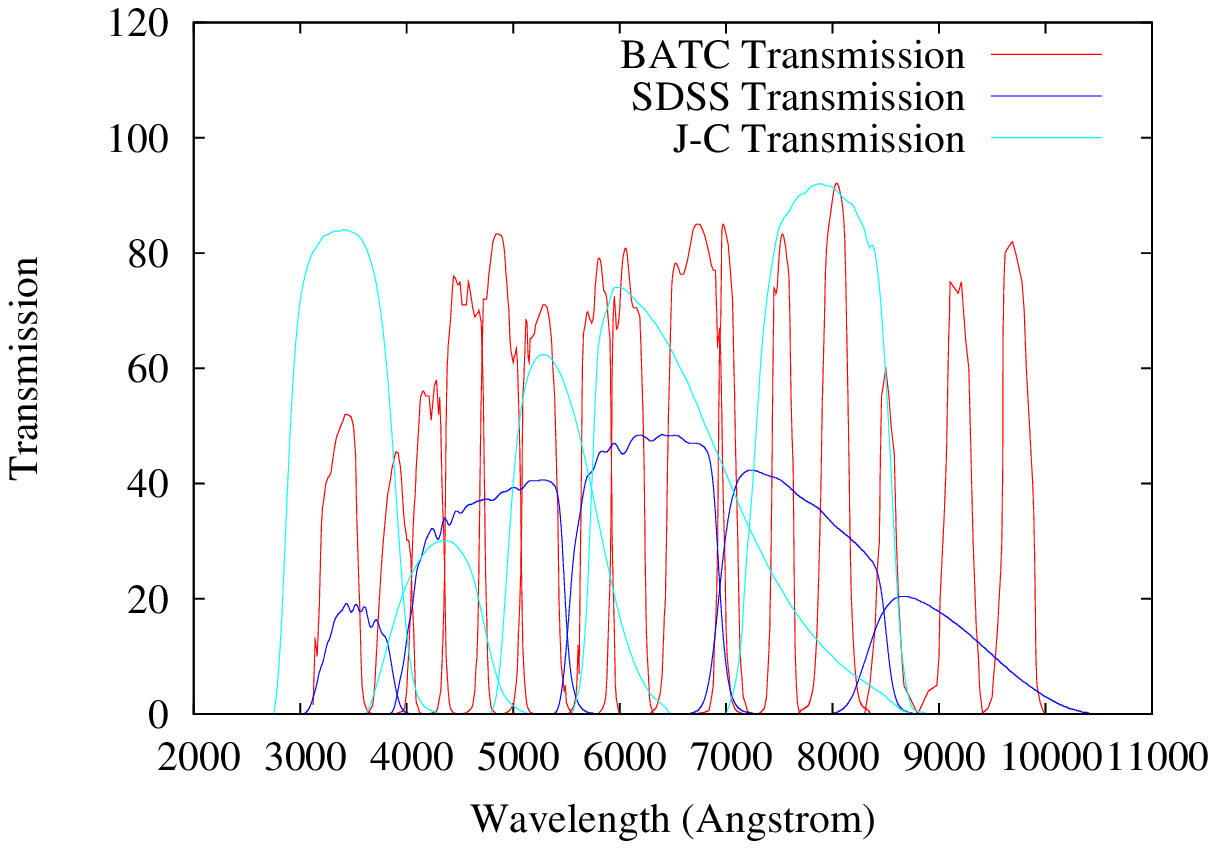}}
\caption{Composite quasar spectra and Transmission curves of three filter
systems.}
\label{F7}
\end{minipage}
\end{figure*}

Several authors have studied the colour or spectral behaviour of 3C~345. They
all found a RWB or steeper-when-brighter behaviour in
the optical domain in this object \citep[e.g.,][]{kidger90b,schramm93,zhang00,
mihov08}. However, there may be a cutoff magnitude ($R\approx15.5$) in the
colour-magnitude correlation, as noted by \citet{mihov08}. When the object
is brighter than that magnitude, its colour is much less dependent on the
magnitude. This cutoff magnitude may also exist in the colour-magnitude
diagrams in \citet{mihov08} and \citet{zhang00}, but is not so obvious as
in \citet{schramm93}.

It has been suggested that a few components in the {\it UV} to
blue band of the spectrum of 3C~345 may be responsible for the RWB or
steeper-when-brighter behaviour \citep{kidger90b,schramm93,mihov08}. These
components include the Mg\,{\sc ii} emission line at 2798 {\AA} and the blue
bump (BB) from 1000 to 4000 {\AA} in the rest frame of this object. The BB
itself consists of a forest of the Fe\,{\sc ii} emission lines, the Balmer
continuum re-emission and the 10\,000--40\,000 K black body emission. Unlike
the underlying non-thermal continuum (hereafter NTC), which is believed to
come from the jet, these
features come either from the accretion disc or from the BLR. Their
variation properties should resemble those of the same emission features of
normal quasars, i.e., they are variable but have smaller amplitudes
and longer time-scales than the NTC variability of blazars
\citep*[e.g.][]{ulrich97,benitez10}. When 3C~345 is in low state, these less
variable emission features (hereafter LVEFs) dominate the fluxes and dilute
the variability at short wavelengths. So the variation amplitudes at short
wavelengths are smaller than those at long wavelengths, as manifested by
\citet{bregman86} and this work (\S3), leading naturally to a RWB behaviour
when the measurements at short wavelengths ($U$, $B$, $V$, $c$ or $e$) are
included in the colour calculation \citep{dai11}. When 3C~345 is in high
state (brighter than the cutoff magnitude of $R=15.5$), the underlying
NTC become strong and may match the LVEFs in strength. Then the object
will show a stable colour, as illustrated in \citet{schramm93}.

On the other hand, when the colour or spectral index is computed by using
two measurements at long wavelengths where the LVEFs are not included in,
we'll get a BWB or flatter-when-brighter behaviour for 3C~345, as for its
BWB behaviour in the $i-m$ and $i-o$ colours in this work and its
flatter-when-brighter trend in the infrared regime in \citet{bregman86}.

Similar colour behaviour has been observed for at least two more FSRQs. In
3C~454.3, the colour is RWB in faint state, reaches a `saturation' at
$R\approx14$, and turns into BWB in bright state \citep{villata06,
raiteri08,sasada10a}. These behaviours were also explained in terms of a
thermal component, probably from the accretion disc, in the emission of
3C~454.3 \citep{raiteri07,raiteri08,sasada10a}. For PKS 1510-089, which has a
pronounced BB in its spectrum \citep{singh97,kataoka08}, it exhibits a RWB
trend except for its prominent flare, which can be explained by the strong
contribution of thermal emission from the accretion disc \citep{sasada10b}.

Therefore, the Mg\,{\sc ii}+BB emission can significantly change the colour
behaviour of FSRQs. Except for the Mg\,{\sc ii} line and the BB, there are
some other strong emission features in the spectrum of FSRQs, such as the
Ly$\alpha\,\lambda1216$, C\,{\sc iv}\,$\lambda1549$, C\,{\sc iii]}$\,\lambda
1909$, H$\beta\,\lambda4861$+[O\,{\sc iii}]\,$\lambda\lambda4959,5007$+Fe\,{\sc
ii} and H$\alpha\,\lambda6563$ lines. These features emanate from the BLR
rather than from the jet, and are thus expected to be less variable than the
underlying NTC. They may also significantly change the colour behaviour
of FSRQs when one or more of them is included in the passbands that are used
in the computation of the colour index, especially when the observations are
made with an intermediate- or narrow-band filter system.

In order to assess how the emission lines change the measured fluxes in the
passbands they reside, we then made a number of simulations by convolving
the FSRQ spectrum with the transmission curves of several filter systems.
The spectrum was shifted from 0 to 3 with a step of 0.01. Each spectrum
was convolved with the transmission curves, so as to see the redshift
effects on the measured fluxes.

We used a composite quasar spectrum to mimic the FSRQ spectrum in order to
have a high signal-to-noise ratio and a large enough wavelength coverage.
The FSRQ spectrum resembles the normal quasar spectrum in the content of the
emission lines. However, their continua may differ much. The quasar spectrum
usually has a strong BB, while the FSRQ spectrum has no or only weak BB due
to the prominence of the strongly beamed jet emission \citep{jolley09}. We
kept the emission features of the composite quasar spectrum but partly
removed the BB from the continuum. The modified composite quasar spectrum
was then used in the convolutions.

The composite quasar spectrum was adopted as a combination of the {\it HST UV}
composite from 650 to 1150 {\AA} \citep{zheng97}, the 2dF quasar composite
from 1150 to 6930 {\AA} \citep{croom02} and a near-infrared composite from
6930 to 10600 {\AA} \citep{glikman06}. We assigned a spectral index of 1.52
\citep{fossati98} to the continuum from 1000 to 10600 {\AA} and a spectral
index of 2.6 for the continuum from 650 to 1000 {\AA}, so as to add a weak BB
into the continuum. The spectral index $\alpha$ is defined as $f_\nu \sim
\nu^{-\alpha}$. This modified composite spectrum is plotted in the left
panel of Fig.~\ref{F7}. For a comparison, the composite of the Large Bright
Quasar Survey \citep[LBQS,][]{francis91} is also plotted, demonstrating the
strong BB peaked at around 1300 {\AA}. The right panel of Fig.~\ref{F7}
displays the transmission curves of the BATC (15 intermediate bands), SDSS
(five broad bands, $u,g,r,i$ and $z$) and Johnson-Cousins (J-C, five broad
band, $U,B,V,R$ and $I$) filter systems.

\begin{figure}
\includegraphics[width=8.4cm]{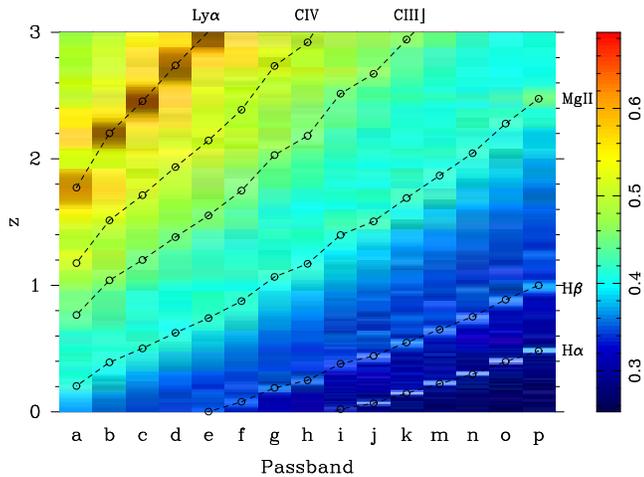}
\caption{Observed fluxes of a FSRQ in the 15 intermediate BATC bands. The
fluxes is plotted with colour in logarithm scale. The open circles denote
the redshifts at which the emission lines move to the weighted centers of
the passbands. The dashed lines are used to guide the 'movements' of the
emission lines with increasing redshift.}
\label{F8}
\end{figure}

The convolutions were made at first between the composite spectrum and the
BATC transmission curves. The resulting fluxes in the 15 intermediate bands
were plotted in logarithm scale as colours in Fig.~\ref{F8}. The open circles
label the redshifts at which the emission lines move to the weighted centers
of the passbands. One can see from that figure that the strong emission
lines move gradually from the short- to long-wavelength BATC passband with
increasing redshift, as shown by the dashed lines. This figure demonstrates
clearly that the strong emission lines can significantly enhance the measured
fluxes in the passbands they are included in and that the enhancement changes
with redshift. At the redshift close to 0.6, the fluxes of 3C~345 at the $c$,
$d$ and $e$ bands are enhanced and dominated by the Mg\,{\sc ii} line and
the Fe\,{\sc ii} lines on its both sides, especially during its faint state.
On the other hand, the fluxes in the $i$, $m$ and $o$ bands are not enhanced
by any strong emission lines and are dominated by the emission from the jet.
So the $e-i$, $e-m$, $c-i$ and $c-o$ colours become RWB, while the $i-m$ and
$i-o$ colours tend BWB, as mentioned at the beginning of this section.

\begin{figure*}
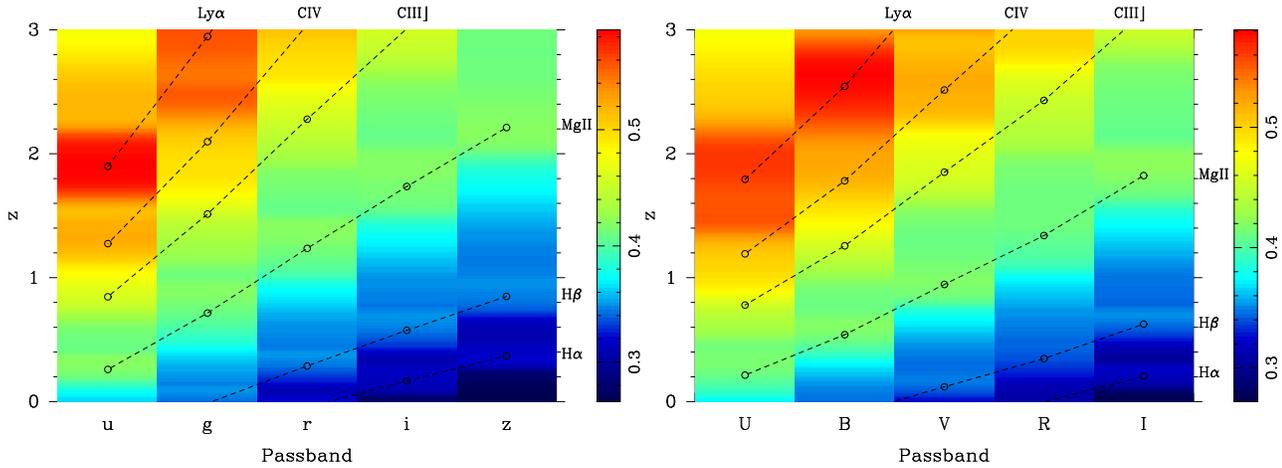

\begin{minipage}{17.6cm}
\mbox{\includegraphics[width=8.4cm]{fig/sdssqsosed.ps}
      \includegraphics[width=8.4cm]{fig/jcqsosed.ps}}
\caption{Same as Fig.~\ref{F8} but for the SDSS and J-C filter systems.}
\label{F9}
\end{minipage}
\end{figure*}

Similar convolutions were made between the composite spectrum and the SDSS
and J-C transmission curves. The results are plotted in Fig.~\ref{F9}. Due
to the much broader passbands, which lower the flux dominance of the emission
lines over the underlying NTC, and the partly overlapping of the
adjacent passbands (see Fig.~\ref{F7}, {\sl right}), the 'movements' of the
emission lines on the two maps are not so manifest as in Fig.~\ref{F8},
but the overall trends can still be seen. At some redshifts, more than one
emission lines can be included in a same passband. For example, both the
Ly$\alpha$ and C\,{\sc iv} lines are included in the SDSS $g$ band and in the
J-C $U$ band at a redshift of about 2.40 and 1.45, respectively, leading to
a considerably enhanced fluxes at those two passbands. At the redshift close
to 0.6, the $B$-band flux of 3C~345 is enhanced by the flux of Mg\,{\sc ii}
line, so its $B-V$ and $B-R$ colours show RWB trends when it is in low state.
In the case of 3C~454.3, which is at a redshift of 0.859, its $V$-band flux
is enhanced by the fluxes of the Mg\,{\sc ii} line. So its $V-R$ and
$V-K_{\rmn s}$ colours show RWB trends when it is in low state.

The simulations clearly illustrate that the strong emssion lines can change
the colour behaviours of FSRQs. Previous explanations for the RWB of FSRQs,
either by the Mg\,{\sc ii}+BB or by a thermal component, have in fact included
the contributions from the strong emission lines of the Ly$\alpha$, C\,{\sc iv}
and C\,{\sc iii}] lines. On the other hand, our simulations focus mainly on
the emission lines, whose equivalent widths are usually less than 80 {\AA} in
the rest frame \citep{peterson97}. For the much broader 'emission' feature,
the BB, which can span from less than 1000 to about 4000 {\AA}, a wavelength
range much broader than the FWHM ($<1500$ {\AA}) of a broad-band filter, the
simulations cannot demonstrate whether it can change the colour behaviours.
However, in the real cases, the broad (the BB) and narrow (the emission
lines) LVEFs usually act simultaneously, especially in the $B$ and $V$ bands
at low redshifts. This will lead to a RWB trend for the $B-R$, $V-R$, $V-J$
and $V-K_{\rmn s}$ colours when the object is in faint state, as in the cases
of 3C~345, 3C~454.3 and PKS~1510-089.

The simulations are not conclusive due to the different strengths of the
LVEFs in different objects. In some objects, the LVEFs are intrinsically
weak. Then the colour behaviour is expected to be BWB. This is in analogy to
the case of BL Lac objects. Also, because of the broad passbands of the SDSS
and J-C filters, there is a good chance for both the blue ($u$, $g$, $U$ and
$B$) and the red passbands ($r$, $i$, $z$, $V$, $R$ and $I$) to include a
strong emission feature. Then whether the object is RWB or BWB will depend
on the interplay between the enhancements of the fluxes in
the blue and red passbands. The most reliable way to study the spectral
behaviour of FSQRs is to fit the continuum of their spectra. This will
eliminate the impact of the LVEFs to the observed fluxes.

Most recently, \citet{gu11} assembled a sample of 29 FSRQs in the SDSS
Stripe 82 region. By fitting a power-law to the SDSS $ugriz$ photometric
data, they found only one FSRQ showing RWB trend. This is quite different
from previous results \citep{fan00,gu06,hu06,rani10}. The fitting of five
photometric data points is somewhat similar to the fitting of the continuum,
because the fluxes of the LVEFs cannot dominate in all five passbands, as
can be seen in Fig.~\ref{F9}. So their results on the colour behaviour of
FSRQs are more reasonable than the previous results.

\section{Conclusions}
We carried out a three-colour monitoring programme on the FSRQ, 3C 345 from
2006 February to 2009 June. There is at least one outburst during this period.
The overall variation amplitude is 2.640 mags in the $i$ band. INOVs was
observed on two nights. The BWB and RWB chromatisms were simultaneously
detected in this object when using different pairs of passbands to compute
the colours. The BWB chromatism is a shared property with the BL Lacs, while
the RWB trend is likely due to two LVEFs, the Mg\,{\sc ii} line and the blue
bump, at short wavelengths.

We made numerical simulations by convolving a composite quasar spectrum with
three filter systems. The results indicate that some other strong but less
variable emission lines, such as the Ly$\alpha$, C\,{\sc iv}, C\,{\sc iii}],
H$\beta$+[O\,{\sc iii}] and H$\alpha$ lines, in the spectrum of FSRQs may also
significantly enhance the measured fluxes of the passbands they are included
in, and hence change the colour behaviour of the object.

The emission of BL Lac objects is dominated by the beamed NTC from the
relativistic jet. However, the emission from FSRQs should be a combination
of the NTC from the jet, the thermal emission from the accretion disc and the
line emission from the BLR. The two latter components are collectively name
as LVEFs in this paper. The beamed NTC from the jet tends to show BWB
chromatism, no matter in BL Lacs or FSRQs. The LVEFs in FSRQs can
significantly change their colour behaviours, especially when they
are in faint state.

The factors that govern the colour behaviour of FSRQs can be summarized as
follows.

\begin{enumerate}
\item The relative strengths of the LVEFs and the NTC. The relative strengths
of the two components can be different for different objects, and are related
to the brightness states of the objects. Strong LVEFs but weak NTC will result
in RWB trend, while weak LVEFs and strong NTC will lead to in BWB chromatism.
\item The redshift. The redshift decides the wavelengths of the observed
LVEFs.
\item The passbands used for colour calculation. Different passbands will
include different LVEFs.
\end{enumerate}

\section{Acknowledgements}
The authors thank the anonymous referee for insightful suggestions and
comments. We thank M. Gu for useful discussions. This work has been supported
by Chinese National Natural Science Foundation grants 10873016, and 11073032,
and by the National Basic Research Programme of China (973 Programme) No.
2007CB815403.

\end{document}